\documentclass[prd,aps,superscriptaddress,twocolumn,floatfix,nofootinbib]{revtex4-1}
\pdfoutput=1

\usepackage{amsfonts}
\usepackage{amsmath}
\usepackage{amssymb}
\usepackage{bm}
\usepackage{dcolumn}
\usepackage{graphicx}   
\usepackage[latin1]{inputenc}
\usepackage{latexsym}
\usepackage{rotating}
\usepackage{hyperref}
\usepackage{graphicx}
\usepackage{color}

\newcommand\be{\begin{equation}}
\newcommand\ba{\begin{eqnarray}}
\newcommand\ee{\end{equation}}
\newcommand\ea{\end{eqnarray}}

\begin{document}

\title{Decay of ALP Condensates via Gravitation-Induced Resonance}

\author{Robert Brandenberger}
\email{rhb@physics.mcgill.ca}
\affiliation{Department of Physics, McGill University, Montr\'{e}al,
  QC, H3A 2T8, Canada}

\author{Vahid Kamali}
\email{vkamali@ipm.ir}
\affiliation{Department of Physics, McGill University, Montr\'{e}al,
  QC, H3A 2T8, Canada}
\affiliation{
  Department of Physics, Bu-Ali Sina (Avicenna) University, Hamedan 65178,
  016016, Iran}
\affiliation{
  School of Physics, Insitute for Research in Fundamental Sciences (IPM),
  19538-33511, Tehran, Iran}
  
\author{Rudnei O. Ramos} 
\email{rudnei@uerj.br}
\affiliation{Departamento de Fisica Teorica, Universidade do Estado do
  Rio de Janeiro, 20550-013 Rio de Janeiro, RJ, Brazil }
\affiliation{Department of Physics, McGill University, Montr\'{e}al,
  QC, H3A 2T8, Canada}
  

\begin{abstract}

Oscillating scalar field condensates induce small amplitude
oscillations of the Hubble parameter which can induce a decay of the
condensate due to a parametric resonance instability~\cite{Mesbah}. We
show that this instability can lead to the decay of the coherence of
the condensate of axion-like particle (ALP) fields during the
radiation phase of standard cosmology for rather generic ALP parameter
values, with possible implications for certain experiments aiming to search
for ALP candidates. As an example, we study the application of this
instability to the QCD axion. We also study the magnitude of the
induced entropy fluctuations.

\end{abstract}

\maketitle

\section{Introduction} 
\label{sec:intro}

Axions and ALPs are amongst the well-motivated candidates for dark
matter (see, e.g., Refs.~\cite{axionrevs,axion} for recent reviews of
axions as dark matter, and Ref.~\cite{ALPrevs} for reviews of
ALPs). It is often assumed that the associated fields are misaligned
relative to the minium of their zero temperature potential in the
early universe and then at some time begin to coherently oscillate
about their ground state values\footnote{Our analysis applies to
  condensates which are oscillating coherently over the entire current
  Hubble radius. Such condensates could have been set up during an
  early phase of cosmological inflation.}.  {}For certain parameter
ranges, these coherently oscillating ALP of axion fields can provide
an important contribution to Dark Matter (DM). As was pointed out
recently in Ref.~\cite{Mesbah} (see also Ref.~\cite{Natalia} for some
earlier related work),  a coherently oscillating scalar field
condensate $\phi$ in the early universe will induce oscillations in
the Hubble expansion parameter superimposed on the overall decreasing
expansion rate.  In a similar way that a coherently oscillating
inflaton condensate will lead to a parametric resonance instability of
the condensate to the production of quanta of any field coupled to the
inflaton~\cite{TB, DK}~\footnote{This effect is known as {\it
    preheating}. See also Refs.~\cite{KLS, STB, KLS2} for other early
  works, and Refs.~\cite{ABCM, Karouby} for reviews.}, in particular
to quanta of the inflaton field itself,  the oscillations in the
Hubble parameter induced by the oscillations of the homogeneous
condensate of $\phi$ will induce resonant production of $\phi$
quanta. This instability can destroy the coherence of the $\phi$
condensate.

In this paper we study the application of this general effect to
proposed dark matter condensate fields, e.g., axions or ALPs,  or more
generally ``wave dark matter''. We find that in a wide range of
parameter space of interest to DM model building,  the instability is
effective and the coherence of the condensate is destroyed. While this
effect does not change the overall energy density of matter (the
energy density of the condensate and of an assembly of individual low momentum
particle quanta both redshift as matter), coherence effects are washed
out. This may have implications for certain searches for wave DM
candidates.

We are interested in condensates which are present in the radiation
phase of standard cosmology.  We work in the context of a homogeneous,
isotropic and spatially flat metric with scale factor $a(t)$, where
$t$ is the physical time (with associated temperature $T$).  We will
be focusing on the radiation phase of Standard Big Bang cosmology when
$a(t) \sim t^{1/2}$. This phase ends at the time $t_{eq}$ of equal
matter and radiation (with associated temperature $T_{eq} \sim
1{\rm{eV}}$. The expansion rate $H(t)$ of space is related to the
temperature via the {}Friedmann equation
\be 
H^2 \sim T^4 m_{pl}^{-2}, 
\ee
where $m_{pl}$ denotes the Planck mass.  One notes that the mechanism
described here is model-independent and applies to any oscillating
condensate field.  The mechanism is also operative in the absence of
nonlinear interactions of $\phi$ quanta.  The instability of axion
condensates in the presence of nonlinearities has been studied in
Ref.~\cite{Guth}, and the analysis generalizes to ALPs, as mentioned
in that same reference.  In the presence of couplings of $\phi$ to
other fields,  parametric resonance instabilities have been analyzed
in various works. In particular, in Ref.~\cite{Wei2}, the standard
coupling of the QCD axion to photons was supplemented with  a coupling
to a dark photon, and the tachyonic instability (closely related to
the parametric resonance instability studied here) of the dark photon
mode equation allowed an opening up of the axion window (see
also~\cite{other}).  In~\cite{Gonzalo} the decay of an ALP condensate
into photons has been analyzed. However, the resonance  is
model-dependent and was found to be of ``narrow resonance'' type, and
its effects hence are much reduced when the expansion of space is
taken into account. The instability which we find is, one the other
hand, of ``broad resonance'' type and hence robust towards taking the
expansion of space into account.

In spirit, our work is related to that of Ref.~\cite{Wei}, where the
transfer of energy from low momentum modes to momenta of the order of
the ALP mass was considered, in that case triggered by the
cosmological fluctuations set up in the primordial universe. There is
also related work on kinetic fragmentation of an ALP
condensate~\cite{Servant}.

This work is organized as follows. In section~\ref{section2}, we
first review the basic effect, following the discussion
in~\cite{Mesbah}. We next (section~\ref{section3}) 
apply the analysis to ALPs and show that, at
least in the class of models which we study, the coherence of the ALP
condensate is destroyed.  The process is robust
and can happen generically in the radiation dominated phase and prior
to the time of matter-radiation equality. In section~\ref{section4}, we
then study in more detail the special case of the QCD axion.  We study
the contribution of the produced axion fluctuations to the entropy in
section~\ref{section5}, and in section~\ref{section6}, we demonstrate
that, although entropy fluctuations on infrared scales are induced by
our effect, they are too small to have an interesting effect on the
amplitude of the curvature fluctuations. {}Finally, in
section~\ref{section7},  we conclude with a discussion of the
implications for experiments designed to search for ALPs and axions.

Throughout this paper, we work with the natural units, in which the
speed of light,  Planck's constant and Boltzmann's constant are all
set to $1$, $c=\hbar=k_B=1$.

\section{The Basic Effect}
\label{section2}

We consider a scalar field $\phi$ with a potential $V(\phi)$ which is
quadratic about its minimum 
\be 
V(\phi)  =  \frac{m^2}{2} \phi^2 \, .  
\ee
The equation of motion of a homogeneous condensate of $\phi$ in an
expanding universe is
\be 
{\ddot{\phi}} + 3 H {\dot{\phi}} + m^2 \phi  =  0 \, .  
\ee
It is convenient to consider the rescaled field $\psi$ defined via
\be 
\psi \equiv  a^{3/2} \phi , 
\ee
whose equation of motion is
\be 
\label{psieq} 
{\ddot{\psi}} +  \left[ m^2 - \frac{3}{4}
  \left(\frac{{\dot{a}}}{a}\right)^2 - \frac{3}{2}
  \frac{{\ddot{a}}}{a}\right]\psi = 0 \, .  
\ee

The condensate $\phi$ will be frozen by Hubble friction until the time
$t_o$ when $H \sim m$. Thereafter, it will oscillate with an amplitude
$A$ which scales like
\be 
A^2 \, \propto \, T^{3} \, .  
\ee
The rescaling of $\phi$ was done such that the amplitude ${\cal{A}}$
of $\psi$ remains constant. We will normalize the scale factor $a(t)$
such that $a(t_o) = 1$. This implies that ${\cal{A}}$ is the amplitude
of $\phi$ at the time $t_o$.

Let us consider some time $t_d$ after the condensate has started to
oscillate.  {}For times after $t_o$, the gravitational terms in
Eq.~(\ref{psieq}) can be neglected if we are interested in processes
that take place on a time scale shorter than the Hubble expansion
time.  Hence, we have that
\be 
\psi(t) \, = \, {\cal{A}} \cos[m(t - t_d)] \, .  
\ee
The oscillations of the condensate will (via the {}Friedmann
equations) induce an oscillating contribution to the scale factor,
which is superimposed on the usual radiation phase scaling.
{}Following~\cite{Mesbah}, we make the ansatz
\be 
\label{ansatz} 
a(t) \, = \, a_0(t) + b(t) \, , 
\ee
where $a_0(t)$ is the usual radiation phase evolution of the scale
factor, and $b(t)$ is a perturbation whose amplitude is obviously
suppressed by the ratio of the energy density of the condensate
divided by the total energy density.  As shown in~\cite{Mesbah}, to
leading order the solution for $b(t)$ is
\ba 
\label{bresult} 
b(t) \, &=& \, \frac{\pi}{3} G m^2 {\cal{A}}^2
(\eta - \eta_d)^2 \nonumber \\ 
& & \, + \pi G m^2 {\cal{A}}^2
\,\frac{\eta - \eta_d}{m}  \cos\left[\frac{1}{2} \frac{m}{t_d} (\eta^2
  - \eta_d^2) \right], 
\ea
where $\eta$ ($\eta_d$) is the conformal time associated with $t$
($t_d$).

As pointed out in~\cite{Mesbah}, the oscillations in $a(t)$ can induce
a gravitational parametric resonance instability in all fields. In
particular, it can excite inhomogeneous modes of $\phi$. If the
resonance is effective, this process will destroy the coherence of
$\phi$. The equation of motion for the {}Fourier modes $\psi_k$ of
$\psi$ is
\be 
\label{modeeq} 
{\ddot{\psi}}_k + \left[ m^2 - \frac{3}{4}
  \left(\frac{{\dot{a}}}{a}\right)^2 - \frac{3}{2}
  \frac{{\ddot{a}}}{a}\right] \psi_k + \frac{k^2}{a^2} \psi_k \, = \,0 \, .  
\ee
Inserting Eqs.~(\ref{ansatz}) and (\ref{bresult}) in
Eq.~(\ref{modeeq})  and neglecting terms with time derivates of $a_0$
(i.e., focusing on the oscillatory term), the mode equation
(\ref{modeeq}) for infrared modes (modes with $k/a < m$)
becomes\footnote{Since the particles which are being produced have
  $k_p < m$ ($k_p$ being the physical momentum),  the instability we
  are discussing will not change the fact that the time averaged
  equation of state of the $\phi$ field is that of cold dark matter.}
\be 
\label{reseq} 
{\ddot{\psi}}_k + \left\{ m^2 - 12 \pi G m^3
    {\cal{A}}^2 (t - t_d)  \cos [ 2 m (t - t_d) ] \right\} \psi_k =  0. 
\ee
This is the equation of motion of a harmonic oscillator with an
oscillating contribution to the mass. Except for the fact that this
contribution to the mass has an amplitude which grows in time, this is
the usual Mathieu equation~\cite{Mathieu}, which has a parametric
resonance instability.  In~\cite{Mesbah} it was shown that this
instability persists also when taking the time dependence of the
amplitude into account\footnote{It would be of interest to provide a
  better analytical analysis of the solutions of
  Eq.~(\ref{reseq}).}. An intuitive way to understand this result is
the following: the time scale of the instability is (for the parameter
range in which the instability is effective) short compared to the
Hubble time scale, the time scale on which the coefficient
varies. Hence, within the instability time scale the variation of the
coefficient is negligible.

In terms of a new time variable $z = mt$,  the mode
equation~(\ref{reseq}) takes the form (where in this equation the
overdot stands for the derivative with respect to $z$)
\be 
{\ddot{\psi}}_k + \left[ 1 - q \cos(2z) \right] \psi_k \, = \, 0,
\ee
with
\be 
q \, \sim \, 12 \pi G {\cal{A}}^2 m t_d \, .  
\ee 
The condition for the parametric resonance instability to be effective is $q > 1$,
i.e., 
\be 
\label{decaycond} 
\frac{{\cal{A}}}{m_{pl}} \, >
\frac{1}{\sqrt{12\pi}} \bigl( m t_d \bigr)^{-1/2} \, .  
\ee 
Since ${\cal{A}}$ is constant, the parametric resonance condition can be
satisfied at later times $t_d$ even if the condition is not satisfied
when the condensate oscillations start. 
 
If the decay condition~(\ref{decaycond}) is satisfied, then all modes
 with $k_p < m$ (where $k_p$ is the physical momentum) will undergo
 exponential amplification with 
\be 
\psi_k \, \sim \, e^{\mu (t - t_d)} , 
\ee 
with $\mu = \sqrt{q} m$.  Note that the modes acquire a high occupation number.
 
The second efficiency condition is the requirement that the
instability is rapid on the Hubble time scale, i.e., 
\be 
\label{effcond2} 
\frac{\mu}{H} \, > \, 1 \, .  
\ee
 
\section{Parametric Instability of ALP Condensates due to Hubble Constant Variations}
\label{section3}

Let us now establish that the mechanism proposed in the previous
section is generic during the radiation dominated regime\footnote{Note
  that in this section, for simplicity we drop order one constants 
  which will not be relevant for our main conclusions.}.  Here we
consider a coherently oscillating scalar field condensate for a field
$\phi$ of mass $m$ and amplitude of oscillation $A(t)$. The
oscillations begin when the Hubble damping in the equation of motion
for $\phi$ becomes negligible. This happens when $H \sim m$.
Recalling that we are considering the radiation phase of standard
cosmology,  the {}Friedmann equation immediately yields
\be 
T_o \,\sim \, \left( m m_{pl} \right)^{1/2}, 
\ee
for the temperature $T_o$ when the oscillations start. Equivalently,
\be  
\label{Tonset} 
\frac{T_o}{T_{eq}} \, \sim \,  \left(\frac{m}{T_{eq}} \right)^{1/2} 
\left( \frac{m_{pl}}{T_{eq}}
\right)^{1/2} \, .  
\ee

 Next we want to determine at what temperature $T_d$ the parametric
 resonance instability sets in.  The condition was given in the
 previous section, by Eq.~(\ref{decaycond}). Note that the amplitude
 ${\cal{A}}$ in that condition is the amplitude $A$ of the oscillation
 of the condensate $\phi$ when the oscillations set in.

Assuming that the amplitude of the oscillation of the condensate is
chosen such that the condensate can provide all of the DM, we have
(from the {}Friedmann equation evaluated at the time $t_{eq}$),
\be 
\label{DMcond} 
A(t_{eq})^2 m^2 \, \sim \, T_{eq}^4 \, .  
\ee
and, hence (using the scaling of $A(t)$ discussed in the previous
section),
\be 
{\cal{A}}^2 \, \sim \, T_o^3 T_{eq} m^{-2}.  
\ee
Inserting this result into Eq.~(\ref{decaycond}), solving for $t_d$
and expressing the result in terms of the temperature $T_d$ (again
making use of the {}Friedmann equation) yields
\be 
\label{Tdecay} 
\frac{T_d}{T_{eq}} \, \simeq \, (12 \pi)^{1/2}
\left( \frac{m}{T_{eq}} \right)^{1/4} \left( \frac{m_{pl}}{T_{eq}}
\right)^{1/4}.  
\ee
Note that in the above equation we have expressed the ratios in terms
of $T_{eq}$ since we want to determine for which values of $m$ the
onset of the instability will be before the time of equal matter and
radiation.

Writing the condensate mass as $m = m_1 {\rm eV}$ (i.e., in units of
eV) and inserting the values of $T_{eq}$ and $m_{pl}$ we obtain
\be 
\label{Tonset2} 
\frac{T_0}{T_{eq}} \, \sim \, m_1^{1/2} 10^{14} ,
\ee
and
\be 
\label{Tdecay2} 
\frac{T_d}{T_{eq}} \, \sim \, m_1^{1/4} 10^{8} \,.  
\ee
The observational lower bound on the condensate mass is $m_1 >
10^{-20}$.  Inserting the value of the lower bound into the two above
equations, we see that for this value of $m$ we have $T_{eq} < T_{d} <
T_{o}$, and the scaling with $m_1$ immediately shows that for all
allowed values of $m_1$ we have
\be 
T_{eq} \, < \, T_d \, < \, T_o \, .  
\ee

We have thus seen that for the allowed parameter range of models
considered here, ALP condensates suffer the parametric resonance
instability well during the radiation period of standard cosmology. It
is easy to verify that the efficiency condition~(\ref{effcond2}) is
also satisfied.

\section{Decay of an Axion Condensate} 
\label{section4}

In this section we consider a special case, namely the QCD axion.  A
key difference compared to the previous analysis is that the potential
for the condensate is not present at all times, but sets in during a
particular phase transition, namely the Peccei-Quinn
symmetry~\cite{PQ} breaking.  Another difference compared to the
analysis of the previous section is that we will not assume that the
axion makes up all of the DM, but we will study the potential
instability of the axion condensate more generally as a function of
the axion parameters.

We will assume that the axion dark matter is produced through the
misalignment mechanism,  where there is a coherent initial
displacement of the axion field.  In this case, at early times when
the Hubble friction is large compared to the axion mass, $H > m_\phi$,
the axion field is overdamped and is frozen at some initial value. Later,
when $H < m_\phi$, the axion field becomes underdamped and
oscillations can begin.  {}For small oscillations, we can approximate
the axion potential as a quadratic potential. Thus, the equation of
state oscillates around $w_\phi = 0$, and the energy density scales as
$\rho_\phi \propto 1/a^3$. This behavior is similar to that of
ordinary matter. This is why misalignment axions can then be
considered  as valid DM candidates.  The misalignment scenario for
axion DM production is expected to happen in the radiation dominated
universe, between the TeV and the QCD scales, depending on the value
of the axion decay constant $f_a$.

The axion equation of motion is (for amplitudes $\phi \lesssim f_a$),
\begin{equation}
\ddot{\phi} + 3 H \dot \phi + m^2_\phi \phi =0,
\label{eom}
\end{equation}
where, for a radiation dominated universe,
\begin{equation}
H^2 = \frac{8\pi}{3 m_{\rm Pl}^2} \rho_{r},\;\;\; \rho_r =
\frac{\pi^2}{30} g_*(T) T^4,
\label{hubble}
\end{equation}
where $g_*(T)$ is the number of relativistic degrees of freedom (DoF)
at the temperature $T$.

The axion mass is a function of the
temperature~\cite{Borsanyi:2016ksw}
\begin{equation} 
m_\phi(T) = \left\{
\begin{array}{ll}
\frac{\sqrt{\chi_0}}{f_a} \left( \frac{T_{QCD}}{T} \right)^n, & {\rm
  for}\;\; T\gtrsim T_{QCD} \\ 5.7 \times 10^{-6} \left(\frac{10^{12}
  {\rm GeV}}{f_a}\right) {\rm eV}, & {\rm for}\;\; T\lesssim T_{QCD},
\end{array}
\right.
\label{mT}
\end{equation}
where $\chi_0 \approx (75.6{\rm MeV})^4$, $n\approx 4.08$ and $T_{QCD}
\approx 153 {\rm MeV}$. The axion mass is approximately constant below
the QCD phase transition, and its value will be denoted by $m_a$.

\begin{center}
\begin{figure}[!htb]
\includegraphics[width=6.4cm]{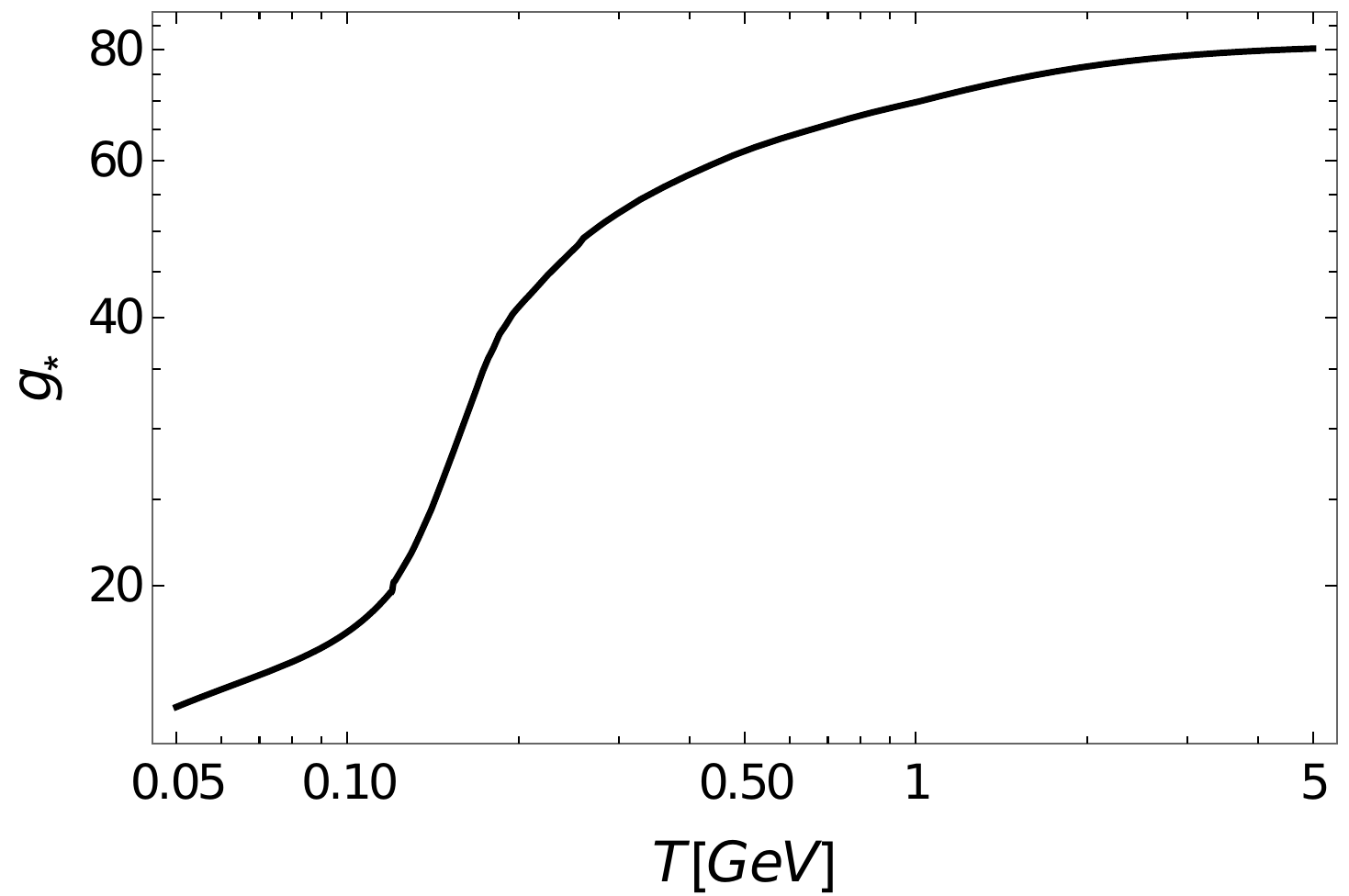}
\caption{The number of relativistic DoF $g_*$ as a function of the
  temperature. An interpolation of the data given in
  Ref.~\cite{Husdal:2016haj} has been used. }
  \label{fig1}
\end{figure}
\end{center}

Taking $m_\phi(T_{o}) = 3 H(T_{o})$, we can determine the temperature
when the axion condensate starts to oscillate coherently. Using
Eqs.~(\ref{hubble}) and (\ref{mT}) (taking $n=4$ in Eq.~(\ref{mT}) for
simplicity), we find
\begin{equation}
T_o \approx 1 {\rm GeV} \left(\frac{10^{12}{\rm
    GeV}}{f_a}\right)^{1/6} \left(\frac{75}{g_*}\right)^{1/12},
\label{Tosc}
\end{equation}
where we have used $g_*\sim 75$ at around the GeV
scale~\cite{Husdal:2016haj}. The number of relativistic DoF $g_*$ is
well known from temperatures about the electroweak scale down to
today. In {}Fig.~\ref{fig1} we show the variation of $g_*$ as a
function of the temperature.

{}For all $T < T_o$ the axion zero mode will satisfy $m_a > 3 H$ and
will be oscillating around the minimum of its potential. Let us
parametrize the initial amplitude of oscillation as
\be 
\phi(T_o) \, = \, c_a f_a \, , 
\ee
where $c_a < 1$ is a positive constant describing the fraction of the
maximal amplitude.  Using the scaling of the oscillation amplitude as
a function of temperature, we can then estimate the contribution of
the oscillating axion condensate to the dark matter density. A careful
analysis~\cite{axionrevs} yields
\be 
\label{fraction} 
\Omega_a h^2 \, \sim \,  10^{-1} c_a^2 \left(
\frac{f_a}{10^{12} {\rm{GeV}}}\right)^{7/6} \, , 
\ee
where $\Omega_a$ is the fractional contribution of axions to the total
energy density, and $h$ is the value of the current Hubble constant in
units of $100 {\rm{km}} s^{-1} {\rm{Mpc}}^{-1}$. The allowed parameter
space for the axion is then determined by the range of values of $c_a$
and $f_a$ for which $\Omega_a \leq \Omega_m$, where $\Omega_m$ is the
total fractional contribution of matter to the energy density budget
of the current universe.

{}From the results of section~\ref{section2} of this paper, we know
that the oscillations of the scale factor that the oscillating
condensate induces lead to parametric resonance growth of long
wavelength axion fluctuations. The resonance is efficient if
\be 
q \, = \, 12 \pi G {\cal{A}}^2 m_a t \, \gg \,1 \, .  
\ee
Recall that ${\cal{A}}$ is independent of time. Hence, the instability
condition is easier to satisfy the later we look.  Inserting the
expression for the axion mass~(\ref{mT}),  making use of the
{}Friedmann equation to express $t$ in terms of the temperature, and
normalizing by $T_{eq}$ yields the efficiency condition on the
temperature when the resonance can effectively set in,
\be 
\frac{T}{T_{eq}} \, < \, 10^5 \left( \frac{f_a}{10^{12}
  {\rm{GeV}}} \right)^{1/2} \, .  
\ee
Thus, we see that for the range $10^8 {\rm{GeV}} < f_a < 10^{12}
{\rm{GeV}}$ of the axion decay constant which is usually considered,
the axion condensate will be unstable towards the resonance effect
considered here.

The axion fluctuations grow exponentially in time $t$ with the rate
\be 
\mu \, \sim \, (12 \pi)^{1/2}  \frac{\cal{A}}{m_{pl}}  (m_a t)^{1/2} m_a \, , 
\ee
where ${\cal{A}}$ is the initial amplitude of oscillation of the
condensate.  We must also check that the instability is efficient on
the time scale of the expansion of space, i.e.,
\be 
\label{axion-instab} 
\frac{\mu}{H} \, \gg \, 1 \, .  
\ee
 Expressing $H$ and $t$ in terms of the temperature $T$, using the
 {}Friedmann equation and normalizing quantities by $T_{eq}$, we
 obtain
\be 
\frac{\mu}{H}(T) \, \sim \,  10^{27} c_a  \left( \frac{10^{12}
  {\rm{GeV}}}{f_a} \right)^{1/2} \left( \frac{T_{eq}}{T} \right)^3 \,.  
\ee
{}From the above we conclude that the axion condensate is rather
generically unstable to the parametric instability which we are
discussing here.

We can now estimate the production of axion quanta due to the
coherent oscillation of the condensate around the minimum of the
potential.  Since it is infrared modes with $k \leq m_a$ which are
excited, the energy density in produced quanta can be estimated to be
(see also~\cite{Mesbah})
\ba 
\rho_{\rm prod}  \, &\sim& \, 4 \pi \int_0^{m_a} dk\, k^4 k^{-1}
e^{2\mu (t-t_0)}  \nonumber \\ 
&\approx& \,  \frac{\pi}{2} m_a^4
e^{2\mu (t-t_0)} \, ,
\label{rhoprod}
\ea
where in the first line we have assumed that the modes begin in their
quantum vacuum state (this gives the $k^{-1}$ factor, while two powers
of $k$ come from the phase space volume, the other two powers come
from the $k^2$ in the energy of an individual mode).
 
%
%
%

The resonant particle production is expected to stop when its
backreaction becomes significant, i.e., when the produced energy
density Eq.~(\ref{rhoprod}) approaches that of the oscillating scalar
field,
\be 
\rho_\phi \, \sim \, \frac{1}{2} m_a^2 \phi^2 \, ,
\ee
which gives for the time interval $\tau$ over which this particle
production mechanism is effective 
\begin{equation}
\tau \, \approx \, \frac{1}{\mu} \ln \left[  \left( \frac{{\cal
      A}}{m_a} \right) \right].
\label{tau}
\end{equation}
Although the argument inside the logarithm in the above equation is
large,
\be 
\frac{\cal{A}}{m_a} \, \sim \, 10^{26} c_a \left(
\frac{f_a}{10^{12} {\rm{GeV}}} \right)^2 \, , 
\ee
its logarithm is of the order $10^2$. Hence, as long as $\mu > 10^2 H$, 
the time scale of backreaction is smaller than the Hubble scale
(which is a condition for our analysis to be self-consistent).

Let us now take a slightly closer look at the efficiency condition for
the resonance, namely that the resonance time scale must be smaller
than the Hubble time.  Since the ratio $\Gamma$ of particle production
is determined by the coefficient $\mu$ in Eq.~(\ref{rhoprod}), or more
precisely, $\Gamma = 2 \mu$, the process will be efficient if $\Gamma> H$.

Recalling again that the window bounds on the  axion decay constant is
typically $10^{8}{\rm GeV } \, \lesssim \, f_a \, \lesssim \,
10^{12}{\rm GeV} $.  {}For these values of $f_a$, from
Eq.~(\ref{Tosc}), then the value for $ T_{o}$ where the axion starts
oscillating is typically $T_{o} \sim 1$ GeV.  {}Note that for $T > T_{\rm QCD}$, the axion mass is determined by the
temperature dependent  expression in Eq.~(\ref{mT}). Then, in the
interval $T_{\rm QCD}\lesssim~T\lesssim~1{\rm GeV}$, we find that
\begin{equation}
\Gamma/H \simeq 2 \times 10^{-3} \left(\frac{10^{12}{\rm GeV}}{f_a}
\right)^{\frac{1}{2}} \left(\frac{70}{g_*(T)} \right)^{\frac{1}{2}}
\left(\frac{1 {\rm GeV}}{T} \right)^{6},
\label{largeT}
\end{equation}
while for temperatures below the QCD scale, the mass of the axion is
determined by its zero temperature term in  Eq.~(\ref{mT}). The ratio
$\Gamma/H$ then now becomes, for $T\lesssim T_{\rm QCD}$,
\begin{equation}
\Gamma/H \simeq 3.6 \left(\frac{10^{12}{\rm GeV}}{f_a}
\right)^{\frac{1}{2}} \left(\frac{17}{g_*(T)} \right)^{\frac{1}{2}}
\left(\frac{153 {\rm MeV}}{T} \right)^{2} \, .
\label{lowT}
\end{equation}
Note again that axion production gets more and more efficient as the
temperature decreases.

\begin{center}
\begin{figure}[!htb]
\includegraphics[width=6.4cm]{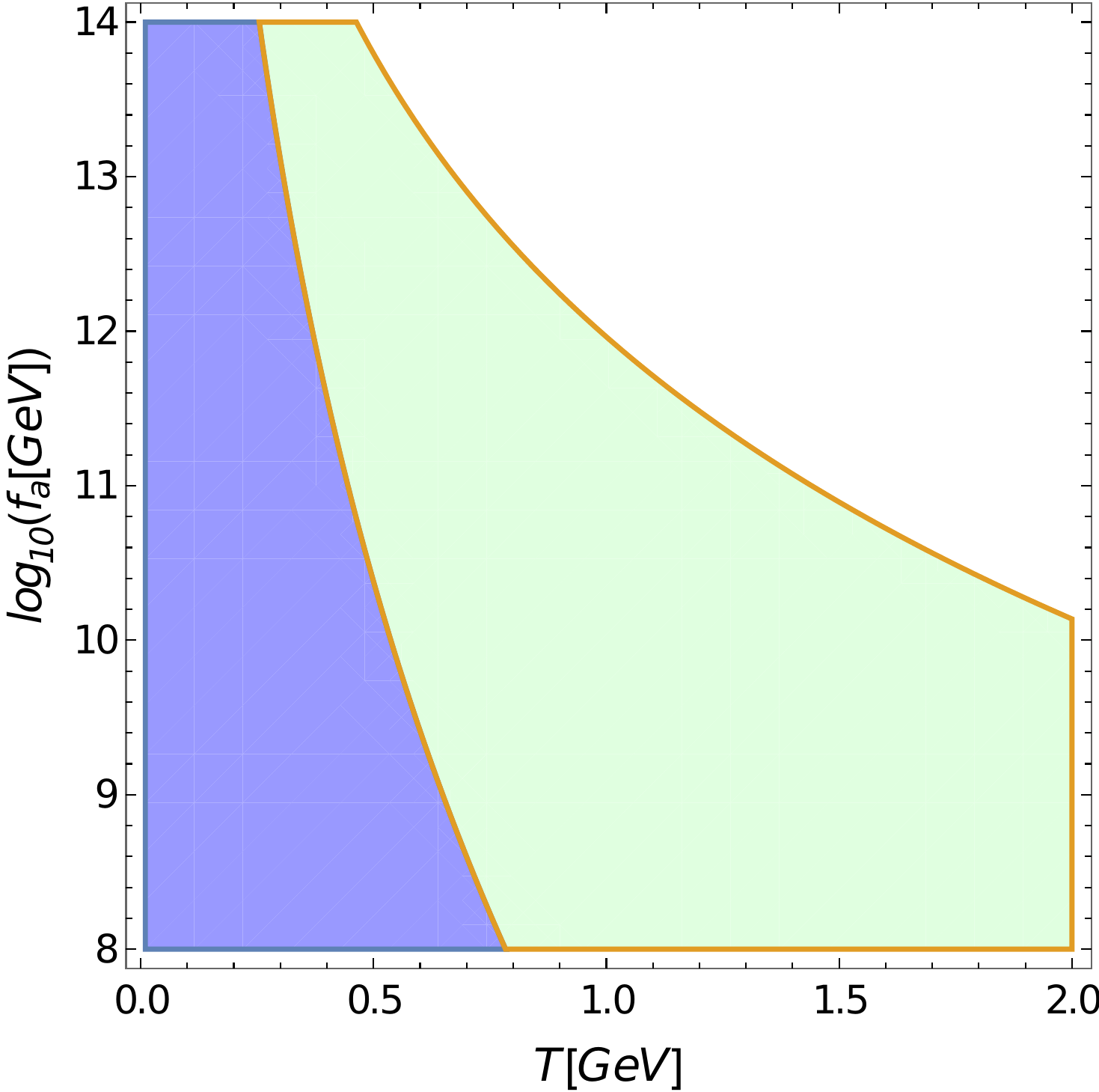}
\caption{The region of parameters for which $\Gamma/H>1$ (blue) and
  for when the axion starts oscillating (green). }
  \label{fig2}
\end{figure}
\end{center}

{}From the data given in {}Fig.~\ref{fig1} and using
Eqs.~(\ref{largeT}) and (\ref{lowT}), we show in  {}Fig.~\ref{fig2}
the region of parameters in terms of the temperature and axion decay
constant where $\Gamma/H>1$ is satisfied. Taking for instance the
temperature $T=100$ MeV, for which $g_*\simeq 17.7$, and for
$f_a=10^{10}$ GeV, we obtain that $\Gamma/H \simeq 8.2\times 10^3$. 

\section{Contribution of Axion Production to the Entropy}
\label{section5}

 In this section we estimate the abundance and entropy of axion DM
 particles produced via our instability mechanism and compare the
 result to the entropy of the thermal bath. The abundance is defined
 by the ratio of the number density of created particles by the
 entropy density,
\begin{equation}
Y_\phi \, = \, \frac{n_\phi}{s}.
\end{equation}
The entropy density of the thermal bath is
\begin{equation}
s_{\rm thermal}(T) \, = \, \frac{2 \pi^2}{45} g_{s,*}(T) T^3,
\label{sT}
\end{equation}
where $ g_{s,*}(T)$ is the number of relativistic DoF for the entropy
at the temperature $T$.

The increase of entropy density for the axion dark matter particles
can be estimated as
\begin{equation}
s_{\rm prod}(t) \, = \, \int \frac{d^3k}{(2 \pi)^3} s_k(t),
\label{intS}
\end{equation}
where $s_k(t)$ is the entropy per mode, which can be defined by the
von Neumann entropy,
\begin{eqnarray}
s_k(t) \, = \, \left[f({\bf k},t) +1\right] \ln  \left[f({\bf k},t) +1
  \right] - f({\bf k},t) \ln  f({\bf k},t).  \nonumber \\
\label{Skt}
\end{eqnarray}
with $f({\bf k},t)$ being the occupation number.

{}For the resonant particle production, we can estimate the occupation
number as
\begin{equation}
f({\bf k},t) \, \sim \, e^{2\mu (t-t_0)}.
\end{equation}
{}For $f({\bf k},t) \gg 1$,
\begin{equation}
s_{\rm prod}(t) \, \sim \, \int \frac{d^3k}{(2 \pi)^3} \ln
\left[f({\bf k},t)  \right] \,  \sim \, \frac{4 \pi m_\phi^3}{3}
\frac{2\mu (t-t_0)}{(2 \pi)^3}.
\label{intS2}
\end{equation}
{}From Eqs.~(\ref{tau}) and (\ref{sT}), we can estimate the ratio
\begin{equation}
\frac{s_{\rm prod} }{s_{\rm thermal}} \sim \frac{30
  \ln\left(\frac{f_a^2}{\pi  m_{\phi }^2}\right) m_{\phi }^3}{\pi  T^3
  g_s(T)}.
\label{ratioS}
\end{equation}
{}For typical values of parameters (for a QCD axion), we have $s_{\rm
  prod} \ll s_{\rm thermal}$.  

It is useful to estimate the ratio (\ref{ratioS}) at the time when the
axion starts oscillating, $m_\phi = 3 H$. This gives a minimum value
for the produced entropy (recalling that particle production will
occur  for some $T < T_{\rm osc}$, see {}Fig.~\ref{fig2}),
\begin{equation} \label{prod}
\frac{s_{\rm prod} }{s_{\rm thermal}}\Bigr|_{T=T_{\rm osc}} \gtrsim
10^2 \ln\left(\frac{f_a^2}{\pi  m_{\phi }^2}\right)
\left(\frac{m_{\phi }}{m_{\rm Pl}}\right)^{3/2}.
\end{equation}
where we have considered $g_{s} \sim g_*$. 

\section{Entropy Fluctuations induced by ALP Condensate Decay}
\label{section6}

It is well known that axion fluctuations can induce primordial entropy
fluctuations~\cite{axion-entropy}.  A similar effect also arises for
ALP fluctuations.  While it is known that there is no parametric
instability of adiabatic fluctuations on super-Hubble scales (e.g.,
induced by the oscillation of the inflaton field at the onset of
reheating (see e.g.~\cite{Fabio1}), it is possible to have
amplification of super-Hubble entropy modes (see, e.g.,~\cite{Fabio2}).
The oscillations of the ALP or axion condensates which we have studied
in this paper induce fluctuations in a sub-dominant matter component,
i.e., entropy fluctuations. In this section, we estimate the magnitude
of the entropy fluctuations induced by the ALP condensate decay
process which we have studied, and compute the induced corrections to
the amplitude of the curvature fluctuations. Since our process
produces high occupation states for infrared modes, one could worry
that the induced curvature fluctuations are too large. 

It is well known that entropy fluctuations induce growing curvature
fluctuations (see, e.g.,~\cite{MFB, RHBfluctsrev} for reviews of the
theory of cosmological fluctuations).  Specifically, a non-adiabatic
pressure fluctuation $\delta P_{\rm{nab}}$ will induce a growth of the
usual curvature fluctuation variable $\zeta$ via the equation
(see~\cite{Wands})
\be 
{\dot{\zeta}}_k \, = \, - \frac{H}{p + \rho} \delta P_{{\rm{nab}},
  k} \, , 
\ee
where the subscript $k$ indicates the comoving momentum mode which we
are considering, and $p$ and $\rho$ are background pressure and energy
density, respectively. 

In our case, the background is the radiation fluid, and the
perturbation is given by the ALP field $\phi$. In this
case\footnote{See also \cite{Hossein} for a more recent application.},
the non-adiabatic pressure fluctuation is~\cite{Wands}
\be 
\delta P_{{\rm{nab}}, k} \, = \, {\dot{p}}\left( \frac{\delta
  p_k}{{\dot{p}}} - \frac{\delta \rho_k}{{\dot{\rho}}} \right) \, ,
\ee
where the fluctuation terms are the $\phi$ pressure and energy density
fluctuations, and the background terms are from the radiation
background. Since $\phi$ is approximately pressureless, we obtain
\be 
\delta P_{{\rm{nab}}, k} \, \simeq \, - \frac{1}{3} \delta \rho_k, 
\ee
and hence,
\be 
{\dot{\zeta}}_k \, \simeq \ \frac{1}{4} \frac{H}{\rho} \delta
\rho_k \, .  
\ee
To estimate the magnitude of the induced curvature fluctuations we
 can integrate this equation from the time $t_d$, when the instability
 sets in (see section~\ref{section3}) until the time $t_d + \tau$ when
 back-reaction shuts off the instability. Since the exponential
 increase in $\delta \rho$ is rapid and terminates on a Hubble time
 scale, we can approximate $\zeta_k$ as
\ba 
\zeta_k \, &\simeq& \, \frac{1}{8} \int_{t_d}^{t_d + \tau} dt
\,\frac{1}{\rho(t) t} \delta \rho_k(t) \, \nonumber \\ 
&\sim& \,\frac{1}{8 t_d} \rho(t_d) \int_{t_d}^{t_d + \tau} dt\, \delta
\rho_k(t_d) e^{2 \mu t} \, \nonumber \\ &\sim& \, \frac{1}{16 \mu t_d}
\frac{\delta \rho_k(t_d + \tau)}{\rho(t_d)} \, .  
\ea
We see from the above equation that the result is suppressed by two
important factors: first the ratio of $H$ to $\mu$, and secondly the
ratio of the density in ALPs to the total radiation density.

The power spectrum $P_{\zeta}(k)$ of the induced curvature
fluctuations on a scale $k$ is, hence, given by the power spectrum of
the ALP density fluctuations on that scale, which is obtained by
integrating $\rho_{\phi}$ from $k = 0$ to $k$ (see
Eq.~\ref{prod}). The result is
\be 
P_{\zeta}(k) \, \sim \, \frac{1}{8} \frac{H(t_d)}{\mu}
\frac{k^4}{m^4} \frac{\rho_{\phi}}{\rho}(t_d) \, .  
\ee
On the far infrared scales relevant to cosmological observations, the
induced power spectrum of curvature fluctuations is, thus, suppressed
by an additional factor of $(k/m)^4$. Hence, we conclude that our
mechanism does not lead to a dangerous amplitude of entropy
fluctuations.

\section{Discussion}
\label{section7}

We have suggested that axion and ALP condensates in the radiation
phase of Standard Big Bang cosmology are unstable to a
model-independent parametric resonance instability, which is triggered
by the contribution of a periodic variation of the effective mass of
the axion or ALP mode functions. This effect comes as  a result of the
periodic variation of the Hubble constant due to the oscillation of
the condensate. We have indicated that the instability is of the
``broad resonance'' type and, hence, robust taking the expansion of
space into account.  Our analysis applies to situations when the condensate is
oscillating coherently on the Hubble scale.

Our effect will destroy the temporal oscillations of axion
and ALP fields coherent over all space.  However,  the modes which
are excited all have momenta smaller than the axion or ALP mass $m$,
are highly excited and hence can be viewed as classical
states (see e.g. \cite{Rodd}) which are oscillating on a time scale of
$m^{-1}$ or larger. Many existing and planned experiments for ALP
detection search for signals involving periodic variations on a time scale 
of $m^{-1}$ (see, e.g., Ref.~\cite{Snowmass} for a recent overview). These
should not be effected by the instability of the global condensate. 
Experiments which, on the
other hand,  search for oscillatory signals which are coherent over
large spatial scales,  may need to be reconsidered  
\footnote{We thank Katelin Schutz and Nick Rodd for 
discussions of these points.}. 

In the case of the QCD axion, the effect does not
appear to have any implications for standard axion detection
experiments that look for interactions of individual axion quanta with
photons. The loss of coherence will, however, impact the suggested
signatures~\cite{Sikivie} of axion DM on galactic scales.

\begin{acknowledgements}

RB wishes to thank the Pauli Center and the Institutes of Theoretical
Physics and of Particle- and Astrophysics of the ETH for
hospitality. The research of RB at McGill is supported in part by
funds from NSERC and from the Canada Research Chair program.   We wish
to thank Katelin Schutz, Nick Rodd and Marco Simonovic for useful
discussions.  V.K. would like to acknowledge the McGill University
Physics Department  for hospitality and partial financial support.
R.O.R. would like to thank the hospitality of the Department of
Physics McGill University.  R.O.R. also acknowledges financial support
of the Coordena\c{c}\~ao de Aperfei\c{c}oamento de Pessoal de
N\'{\i}vel Superior (CAPES) - Finance Code 001 and by research grants
from Conselho Nacional de Desenvolvimento Cient\'{\i}fico e
Tecnol\'ogico (CNPq), Grant No. 307286/2021-5, and from Funda\c{c}\~ao
Carlos Chagas Filho de Amparo \`a Pesquisa do Estado do Rio de Janeiro
(FAPERJ), Grant No. E-26/201.150/2021. 

\end{acknowledgements}


\end{document}